\documentclass[twocolumn,prl,superscriptaddress]{revtex4}
\usepackage{amsmath,amssymb,}
\usepackage{mathrsfs}
\usepackage{graphicx}% Include figure files
\usepackage{dcolumn}% Align table columns on decimal point
\usepackage{bm}% bold math
\usepackage[colorlinks,citecolor=red]{hyperref}

\usepackage{rotating}
\usepackage{floatrow}

\usepackage{lipsum}

\usepackage{bbm}

\begin{document}

\title{Floquet Amorphous Topological Orders in a One-dimensional Rydberg Glass}

\author{Peng He}
\email{penghe@hku.hk}
\affiliation{Department of Physics and HK Institute of Quantum Science \& Technology, The University of Hong Kong, Pokfulam Road, Hong Kong, China}
\affiliation{Hong Kong Branch for Quantum Science Center of Guangdong-Hong Kong-Macau Great Bay Area, Shenzhen, China}

\author{Jing-Xin Liu}

\affiliation{National Laboratory of Solid State Microstructures, School of Physics,
and Collaborative Innovation Center of Advanced Microstructures, Nanjing University, Nanjing 210093, China}

\author{Hong Wu}
\email{wuh@cqupt.edu.cn}
\affiliation{School of Science, and Institute for Advanced Sciences, Chongqing University of Posts and Telecommunications, Chongqing 400065, China}

\author{Z. D. Wang}
\email{zwang@hku.hk}
\affiliation{Department of Physics and HK Institute of Quantum Science \& Technology, The University of Hong Kong, Pokfulam Road, Hong Kong, China}
\affiliation{Hong Kong Branch for Quantum Science Center of Guangdong-Hong Kong-Macau Great Bay Area, Shenzhen, China}

%\affiliation{National Laboratory of Solid State Microstructures, Collaborative Innovation Center of Advanced
%Microstructures, Nanjing University, Nanjing 210093, China}

\date{\today}

\begin{abstract}
\noindent $\textbf{Abstract}$  The topological orders in amorphous systems that lack crystalline symmetry have gained considerable attention recently. Here we propose the Floquet amorphous topological matter, among which the topological orders are explored in experimentally accessible one-dimensional array of randomly pointed Rydberg atoms with periodic driving. The topological properties are comprehensively characterized, considering both the single-particle and many-body perspectives. It is found that the periodic driving leads to rich topological phases of matter. At the single-particle level, we evaluate the real space winding numbers and polarization, revealing robust amorphous topological phases with 0-type and $\pi$-type edge modes. We show a structural disorder induced topological phase transition associated with localization transition in the nonequilibrium system. Remarkably, in the many-body case it is discovered that the amorphous topological order exists in the chain of hardcore bosons, captured by the topological entanglement entropy and the string order. Moreover, feasible experimental probe protocols are also elaborated.
\end{abstract}
%\pacs{42.50.Pq, 37.30.+i, 03.67.Bg, 76.30.Mi}

\maketitle

\noindent $\textbf{Introduction}$

Symmetry-protected topological (SPT) orders have been widely explored in systems with underlying spatial order, such as topological insulators and superconductors \cite{Hasan2010,XLQi2011,Chiu2016,DWZhang2018}, Dirac and Weyl semimetals \cite{Vishwanath2018,YXu2019}. Nevertheless, it has been noted that the inclusion of the spatiotemporal engineering leads to richer topological phases. Even in systems without local crystalline symmetries, including topological quasicystals \cite{Aubry1980,Harper1955,Sarma1990,Sarma2010}, amorphous solids and artificial materials with completely random sites \cite{Corbae2023,Agarwala2017,Yang2019,JHWang2021,Ivaki2020,Sahlberg2020,Hannukainen2022,CWang2022,XCheng2023}, the SPT order can exist. Such topological glassy matter does not rely on the microscopic details of the spatial configuration, manifesting its robustness and facilitating the material fabrications. On another note, with an additional degree of
freedom in the time domain, periodically driven nonequilibrium systems have extended the SPT phases to a new classification scheme, supporting anomalous bulk-boundary relations without equilibrium counterpart \cite{Kitagawa2010,Rudner2013,Goldman2014,Eckardt2017,Roy2017,SYao2017,WWHo2023,TNag2021,Ghosh2023}. Aforementioned findings prompt a search for the new class of SPT phases with the synergy of Floquet topology and amorphous order. Meanwhile a feasible proposal on available  quantum systems is in great need.

Although the studies of SPT orders usually focus on the fermionic systems, they can also naturally appear in bosonic systems when strong interactions are considered and especially when the hardcore condition is fulfilled \cite{Haldane1983,Pollmann2010,XChen2011,XChen2012}. The characterization and classification of the bosonic SPT phases are based on the ground state of the many-body systems. The concept of SPT orders has been extended to the nonequilibrium setup in terms of the Floquet quantum matter \cite{Khemani2016,Pootter2016,Po2016,Potirniche2017,Harper2017,FMei2020,XZhang2022}. For the experimental aspect, the Rydberg atoms combined with Floquet engineering provide a versatile platform for the quantum simulation of many-body physics and the topological theory, due to its high tunability \cite{Saffman2010,Browaeys2020,DWZhang2018,Geier2021,Scholl2022,Kalinowski2023,TFPoon2024,YCheng2024}. The Rydberg atoms are individually controlled by the optical tweezer \cite{Endres2016,Kim2016,Barredo2016,Barredo2018}, and the coupling is tunable via dipole-dipole and van der Waals interactions \cite{Browaeys2016,Orioli2018,Signoles2021}, which allows rather flexible and local design of the lattice configuration. Among various experimental progress \cite{deleseleuc2019,Lienhard2020,Bluvstein2021,Semeghini2021,CChen2023,Juliu2024}, a bosonic SPT phase has been observed with a one-dimensional arrays of $\rm ^{87} Rb$ atoms \cite{deleseleuc2019}. A later theoretical work has shown the existence of the bosonic SPT order in amorphous systems \cite{KLi2021}.

In this paper, we study the SPT phases in a one-dimensional (1D) Floquet amorphous bosonic Rydberg atomic array, based on a experimentally feasible setup. We show the existence of rich SPT phases in both the single-particle level and many-body level. In the single-particle level, we map out the phase diagram according to the real space winding number and the polarization. Topological 0 phase and $\pi/T$ phase can exist in amorphous lattices with suitable design of the periodic drivings. Specifically, we further identify the structural disorder driven phase transitions in our model. We find clear numerical signatures of the Anderson localization in the topological regimes, implying a special gapless Floquet phase. In the many-body level, we consider the hard core conditions at half-filling. We characterize the topology in terms of the topological entanglement entropy (TEE) and the string order, using exact diagonalization (ED) and density matrix renormalization group (DMRG). Furthermore, we provide methods to detect the SPT phases in experiments, with either microwave spectroscopy or the edge fidelity, which extracts the nontrivial boundary physics related with the topology.
%\section{Model Hamiltonian}\label{sec2}

\begin{figure}[htbp]
	\centering
	\includegraphics[width=\textwidth]{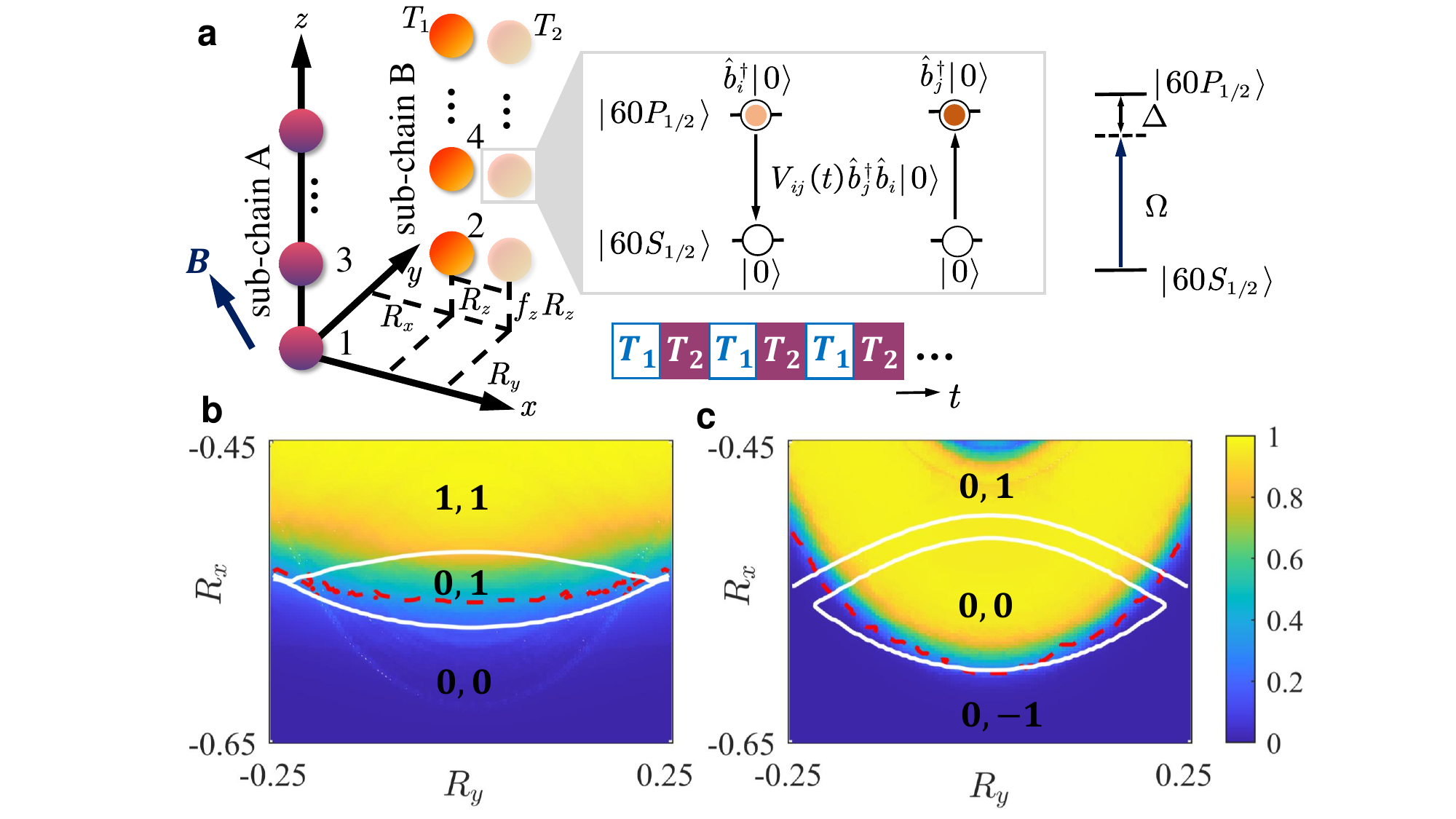}
	\caption{\textbf{Floquet array of Rydberg atoms and the topological phase diagrams.} \textbf{a} Schematics of the driving protocol on the array of Rydberg atoms with atoms at site $2i-1$ and $2i$ forming
a unit cell. The two atoms in a unit cell are initially sperated by a distance $\mathbf{R}=(R_x, R_y,R_z)$, and subject to a periodic modulation. The dipolar exchange interaction between two atoms provides hopping terms. The magnetic field $\mathbf{B}=(-\sqrt{2},0,1)B$ is applied in the $x$-$z$ plane with the polar angle $\theta_m=\arccos(1/\sqrt{3})$ to cancel the hoppings within each subchain. The system is probed by a microwave field with the Rabi frequency $\Omega$ and detuning $\Delta$.  \textbf{b} Phase diagram characterized by $W_0$ for the disordered system, with driving time duration $T_1=T_2=0.4$, driving parameter $\boldsymbol{f}=(0.9,1,1)$, initial distance $R_z=0.72$, structural disorder strength $d=0.8$, and system length $2L=200$. $W_\pi=0$ for this driving condition and disorder strength.  \textbf{c} Phase diagram characterized by $W_\pi$ for the disordered system, with driving time duration $T_1=T_2=0.4$, driving parameter $\boldsymbol{f}=(1,1,0.55)$, initial distance $R_z=0.8$, structural disorder strength $d=0.8$, and system length  $2L=200$. $W_0=0$ for this driving condition and disorder strength. The red dashed lines show the phase boundaries determined by the polarization. The white solid lines show the phase boundaries of a regular system (without the structural disorder) for comparison. The numbers label the values of the  winding numbers $W_0$, $W_\pi$ for regular systems in different phases separated by the white lines for a regular system. The color bar shows the values of the winding numbers for disordered systems.  All the quantities are averaged over 30 random configurations.}
	\label{fig1}
\end{figure}

\noindent $\textbf{Results}$

\noindent \textbf{Model Hamiltonian.}
We consider an array of $2L$ individually trapped $\rm ^{87} Rb$ atoms in a dimerized configuration, as shown in Fig. \ref{fig1}a. For each atom, only two Rydberg states from the $60S_{1/2}$ and the $60P_{1/2}$ manifolds involve, which naturally serve as two distinct hard-core bosonic degrees of freedom. An s-level $ |60 S_{1 / 2}, m_J=1 / 2 \rangle $ corresponds to the ``vacuum" of the many-body system, and a p-level $ |60 P_{1 / 2}, m_J=-1 / 2 \rangle $ corresponds to one occupied boson. Because of the excitation transfer between two Rydberg atoms induced by the dipole-dipole interaction, the system may be described by the following Hamiltonian,
\begin{equation}
\hat{H}(t)=\sum_{i<j}^{2 L} V_{i j}(t) (\hat{b}_i^{\dagger} \hat{b}_j+\hat{b}_j^{\dagger} \hat{b}_i ),\label{eq_ham}
\end{equation}
where $\hat{b}_i^\dagger$ ($\hat{b}_i$) creates (annihilates) a hard-core boson at site i, and $V_{i j}=\tilde{d}_0^2 (1-3 \cos ^2 \theta_{i j} ) / R_{i j}^3$ is the dipolar coupling strength. Here $\tilde{d}_0$ is the the dipole moment between the two subelevels, $R_{ij}$ is the separation between the atoms at site $i$ and site $j$, and  its angle with respect to the magnetic field $\mathbf{B}$ determines $\theta_{ij}$. Specifically, the two atoms in each unit cell is separated by a vector $\mathbf{R}=(R_x,R_y,R_z)a_0$, and subejct to an aditional random dispalcement $R_z^{2i-1}=(i-1+\delta z_i)a_0$ and  $(R_z^{2i}=i-1+R_z+\delta z_i)a_0$, where $\delta z_i$ is the structural disorder uniformly sampled in the range $[-d/2,d/2]$, amorphously shaped the lattice geometry, and $a_0$ is the lattice unit. Furthermore, the atoms in two subchains are aligned along the so-called  ``magic angle" $\theta_{i i+2}=\theta_m\equiv \arccos(1/\sqrt{3})$, so that the coupling within each subchain is vanishing and the chiral (sublattice) symmetry is guaranteed. In following discussions, we set $a_0=1$ and $\tilde{d}_0^2/a_0^3$ as the units of length and energy, respectively, and set $\hbar=1$.

To study the Floquet phases in this system, we consider the stroboscopic modulation of the atom displacement by periodically shift one of the subchain as,
\begin{equation}
\mathbf{R}(t)=\left\{\begin{array}{l}
\mathbf{R},~~t \in\left[m T, m T+T_1\right) \\
\boldsymbol{f} \cdot \mathbf{R},~~t \in\left[m T+T_{1,}(m+1) T\right)
\end{array}, ~m \in \mathbb{Z},\right.
\end{equation}
where $\boldsymbol{f}=(f_x, f_y, f_z)$ is a set of real coefficients. The implementation of the Floquet protocol requires  dynamically toggling of the reconfigurable Rydberg atom arrays. This approach is scalable, with long-time coherence, and has been realized in recent experiments \cite{Bluvstein2022,Kalinowski2023}. Compared to many previous
proposals of Floquet engineering usually demonstrated with a time-dependent detuning or applying an external field, our protocol preserves the system symmetry as described below. The dynamics under driving is described by the unitary evolution $U_T=e^{-i \hat{H}_2 T_2} e^{-i \hat{H}_1 T_1}$, where we denote the Hamiltonian in the
respective time duration $T_1$ and $T_2$ as $\hat{H}_1$ and $\hat{H}_2$ ($T_2\equiv T-T_1$). Then we have an effective Hamiltonian $\hat{H}_T=\frac{i}{T} \ln U_T$. The spectra $\varepsilon_n$ of $\hat{H}_T$ are known as quasienergies and we take $\varepsilon_n\in[-\pi/T,\pi/T]$.

We remark that the Hamiltonian $\hat{H}(t)$ respects a dihedral $\mathbb{Z}_2\times \mathbb{Z}$ symmetry, represented by an anti-unitary operator $\mathcal{S}_{\mathrm{B}}=\prod_{i=1}^{2L} [b_i+b_i^{\dagger} ] K$ and discrete translations in time ($K$ denotes the complex conjugation).The symmetry is exact since $[\hat{H}(t),\mathcal{S}_{\mathrm{B}}]=0$ holds for every individual sample configuration. However, $\hat{H}_F$ does not inherit the symmetry due to $[\hat{H}_1,\hat{H}_2]\neq 0$. After noting that $ U_T$ has a Floquet guage degree of freedom, we can apply two similarity transformations $\hat{G}_j=e^{i(-1)^j \hat{H}_j T_j / 2}$, converting $U_T$ into $\tilde{U}_{T,1}=e^{-i \hat{H}_1 T_1/2}e^{-i \hat{H}_2 T_2}e^{-i \hat{H}_1 T_1/2}$ and $\tilde{U}_{T,2}=e^{-i \hat{H}_2 T_2/2}e^{-i \hat{H}_1 T_1}e^{-i \hat{H}_2 T_2/2}$ respectively, from which we define $\hat{H}_{F,i}=\frac{i}{T}\ln \tilde{U}_{T,i}$ ($i=1,2$). $\hat{H}_{F,i}$ share the same
quasienergies as $\hat{H}_T$, but recover the symmetry of $\hat{H}(t)$.

Our model also preserves a U(1) symmetry $[\hat{H}(t),\hat{N}]=0$ with $\hat{N}=\sum_i \hat{b}_i^\dagger\hat{b}_i$ being the total particle number operator thus $\hat N$ is conserved. In the experiment, a weak global microwave field is applied to create a state with certain excitation numbers if the detuning matches the system energy \cite{deleseleuc2019}. Therefore, we can study this system at both single-particle level and many-body level.

%\section{Single-particle case}\label{sec3}
\begin{figure}[htbp]
	\centering
	\includegraphics[width=\textwidth]{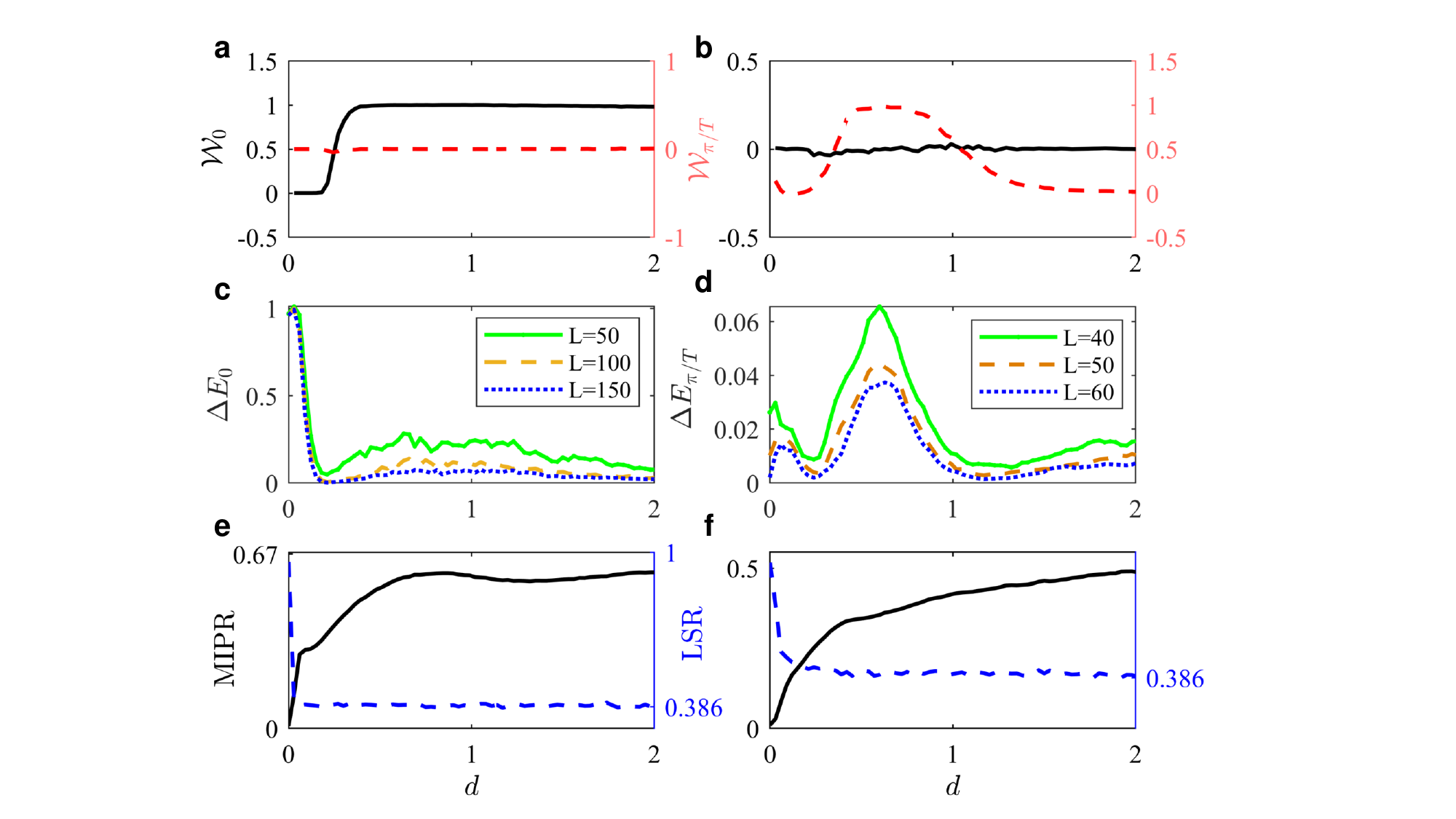}
	\caption{\textbf{The topological phase transitions driven by structural disorder.} The winding number  \textbf{a},  \textbf{b}, quasienergy gap  \textbf{c}, \textbf{d}, and the mean inverse participation ratio and level-spacing ratio \textbf{e}, \textbf{f} versus structrual disorder $d$. We use initial distance $\mathbf{R}=(-0.5,0,0.8)$, driving parameter $\boldsymbol{f}=(0.9,1,1)$, driving time duration $T_1=T_2=0.4$ for \textbf{a}, \textbf{c} and \textbf{e}; and initial distance $\mathbf{R}=(-0.5,0,0.8)$, driving parameter $\boldsymbol{f}=(1,1,0.55)$, driving time duration $T_1=T_2=0.4$ for \textbf{b}, \textbf{d} and \textbf{f}. All the quantities are averaged over 50 random configurations. }
	\label{fig2}
\end{figure}

\noindent \textbf{Single-particle case.}
We first study the topological properties of the single-excitation state manifold of the atom chain. We obtain a single-particle Hamiltonian $\hat{H}_S(t)$ with $[H^S]_{i j}=V_{i j}(t) (1-\delta_{i j})~(1 \leq i, j \leq 2 L)$, under a basis $\hat{b}\equiv \{\hat{b}_1^{\dagger}, \hat{b}_2^{\dagger}, \ldots, \hat{b}_{2 L}^{\dagger} \}$. $\hat{H}_S(t)$ is chiral symmetric $S H^S S^{-1}=-H^S$ with $S=\mathrm{diag}\{(-1)^{j-1}\}_1^{2L}$, thus is equivalent to a Floquet amorphous Su-Schrieffer-Heeger (SSH) model. The nontrivial Floquet topology manifests in the existence of zero-energy edge modes and $\pi/T$-energy edge modes, in which the latter has no equilibrium counterpart.

In the presence of the structural disorder, the quasimomentum is no longer a good quantum number. To characterize the topology of single particle bands, we calculate the real-space winding numbers \cite{Kitaev2006,Bianco2011},
\begin{equation}
\mathcal{W}_{j=1,2}=\frac{1}{2 L^{\prime}} \operatorname{Tr}^{\prime} (S Q_j [Q_j, X] )\,,
\end{equation}
where $X$ is the coordinate operator, $Q_j=\sum_n (|n_j\rangle\langle n_j|-S|n_j\rangle\langle n_j|S^\dagger)$ with $\hat{H}_{F,j}|n_j\rangle=\varepsilon_{j,n}|n_j\rangle$, and $\rm{Tr}'$ denotes the trace over the bulk sites with length $L'=L-2\ell$. The number of 0- and $\pi/T$ -mode edge states relates to $\mathcal{W}_j$ as \cite{Asboth2014,LLin2021,LZhou2018,HWu2020}
\begin{equation}
\mathcal{W}_0=(\mathcal{W}_1+\mathcal{W}_2)/2,\quad \mathcal{W}_{\pi/T}=(\mathcal{W}_1-\mathcal{W}_2)/2.
\end{equation}
The topological phases also carries a nontrivial polarization which is quantized by the chiral symmetry. The polarization is given by,
\begin{equation}
P=[\frac{1}{2 \pi} \operatorname{Im} \ln \operatorname{det} \mathcal{U}-\sum_{l, l^{\prime}, s, s^{\prime}} \frac{X_{l s, l^{\prime} s^{\prime}}}{2 L} ] \quad \bmod 1,
\end{equation}
where the elements of $\mathcal{U}$ read $\mathcal{U}_{mn}\equiv \langle m|e^{ i2\pi X/L}|n\rangle$, and $X_{l s, l^{\prime} s^{\prime}}=r_l\delta_{ll^\prime}\delta_{ss^\prime}$, with $r_l$ being the position of the $l$-th cell and $s,~s'=A,~B$ being the sublattice indices \cite{Resta1998}. Due to its $\mathbb{Z}_2$ nature, the polarization can only distinguishes between even or odd number of pairs of edge states.

Before proceeding, we briefly review the static case, which can be found in Ref.\cite{KLi2021}. The phase diagram is dominated by the interplay between the nearest-neighbor intracell hopping and intercell hopping, whereas the long-range hopping enlarges the regime of topological phases. Similar to the SSH model, a topological limit can be found where the intracell hopping vanishes, given by $\frac{R_y^2-R_x^2+2\sqrt{2}R_xR_z}{(R_x^2+R_y^2+R_z^2)^{5/2}}=0$. A robust topological phase is observed around the topological limit line.

The Floquet driving results in richer phase structures. We map out two typical phase diagrams in the $R_x$-$R_y$ plane according to the winding number and polarization, see Fig. \ref{fig1}b and \ref{fig1}c. Numerical results clearly show the existence of the 0-type ($\pi/T$-type) amorphous SPT order with $ \mathcal{W}_0\approx1$ ($ \mathcal{W}_{\pi/T}\approx1$) according to different drving conditions. Typically, we choose the periodic driving as $\boldsymbol{f}=(0.9,1,1)$ to induce the 0-phase, while $\boldsymbol{f}=(1,1,0.55)$ to induce the $\pi/T$-phase, although in general the type of the Floquet SPT phase does not rely on the driving of certain components of $\mathbf{R}$ (for more discussions, see Supplementary Note 1). Intuitively, as the driving period is increased such that the driving frequency is comparable to the band width, the $\pi/T$-gap would close and then reopen, which makes it possible to find the nontrivial $\pi/T$ edge modes. In both nontrivial phases, the polarization is quantized to $\approx 0.5$, and showing the same phase boundaries with that predicted by the winding numbers. We note that the Floquet system can be topological even when both $\hat H_1$ and $\hat H_2$ are trivial (see Fig. S1(b), Supplementary Note 1). These results reveal the unique features of our model as a nonequilibrium system. Furthermore, we observe a clear deviation of the phase boundary compared to that in the regular limit. Part of the phase is trivial for regular systems while becomes topological under structural disorder, which implies an amorphousness-driven phase transition in the Floquet system. We note that the conclusions are not limited for the stroboscopic driving. We also show that the SPT phase exists under hamornic driving in Supplementary Note 2.

Figure \ref{fig2} illustrates the typical 0-phase and $\pi/T$-phase transition event driven by the disorder. The phase transition is accompanied by gap closure. As shown in Fig. \ref{fig2}c and \ref{fig2}d, the gap quickly drops as the disorder increases. We see that the energy gap decreases as the system size increases, indicating the gap closes in the thermodynamical limit and system enters an ungapped localization phase in the topological regime. The topology is protected by the mobility subgap rather than the spectral gap \cite{YYZhang2012,Loring2015,Cerjan2022,MRen2024}. In such phase all states are localized. The localization properties are identified by both the  level-spacing statistics and the mean inverse participation ratio (MIPR). For level statistics, we use the adjacent level-spacing ratio (LSR): $r(\varepsilon)= [ (1 / (N_\varepsilon-2 )) \sum_i \min  (\delta_i, \delta_{i+1} ) /\max (\delta_i, \delta_{i+1} )]$, where $\delta_i=\varepsilon_i-\varepsilon_{i-1}$ with eigen-quasienergies $\varepsilon_i$'s sorted in an ascending order, and $N_\varepsilon$ is the number of the energy levels counted. As a general rule of thumb, for localized states, $r\approx 0.386$, fulfilling the Poisson statistics, whereas for extended states, $r\approx 0.6$ associated with the Gaussian orthogonal ensemble. On another note, the MIPR is defined by $I=(1/N_E)\sum_{n,x}|\Psi_{n,x}|^4$ with $|\Psi_{n,x}|^2$ being the local density of the n-th state on site $x$ and $N_E$ being the energy levels counted. A state is extended when $I\sim 1/L$. As comfirmed in Fig. \ref{fig2}e and \ref{fig2}f, our model supports robust ungapped 0-type and $\pi$-type localized SPT phase in a wide range of parameters, which is reminiscent of the Floquet Anderson insulator reported in Refs. \cite{Titum2015,Titum2016}. However, the fully localized topological Floquet band in Ref. \cite{Titum2016} relies on the anomalous topology with zero bulk invariant. In contrast, our model supports fully localized topological Floquet band in both normal and anomalous phases, guaranteed that the 1D chiral system is Wannier localizable. We also find that the system could be topological even when the atoms are completely randomly positioned as shown in Fig. \ref{fig2}a. In Fig \ref{fig2}b, the system undergoes a transition from the topological phase to trivial Anderson phase under strong disorder. The two phases are separated by a delocalized critical point, showing that they can not be continuously connected (for more details, see Supplementary Note 3).
%\textcolor{red}{Importantly, in the topological Anderson regime, the bulk is only localized in average .  There must be a delocalized bulk state at some energies due to its topological nature. Therefore, it is possible to induce a delocalized topological bulk mode from both fully localized $\hat{H}_1$ and $\hat{H}_2$ via periodic driving, showing the nonequilibrium aspect of a Floquet amorphous system. }

\begin{figure}[htbp]
	\centering
	\includegraphics[width=\textwidth]{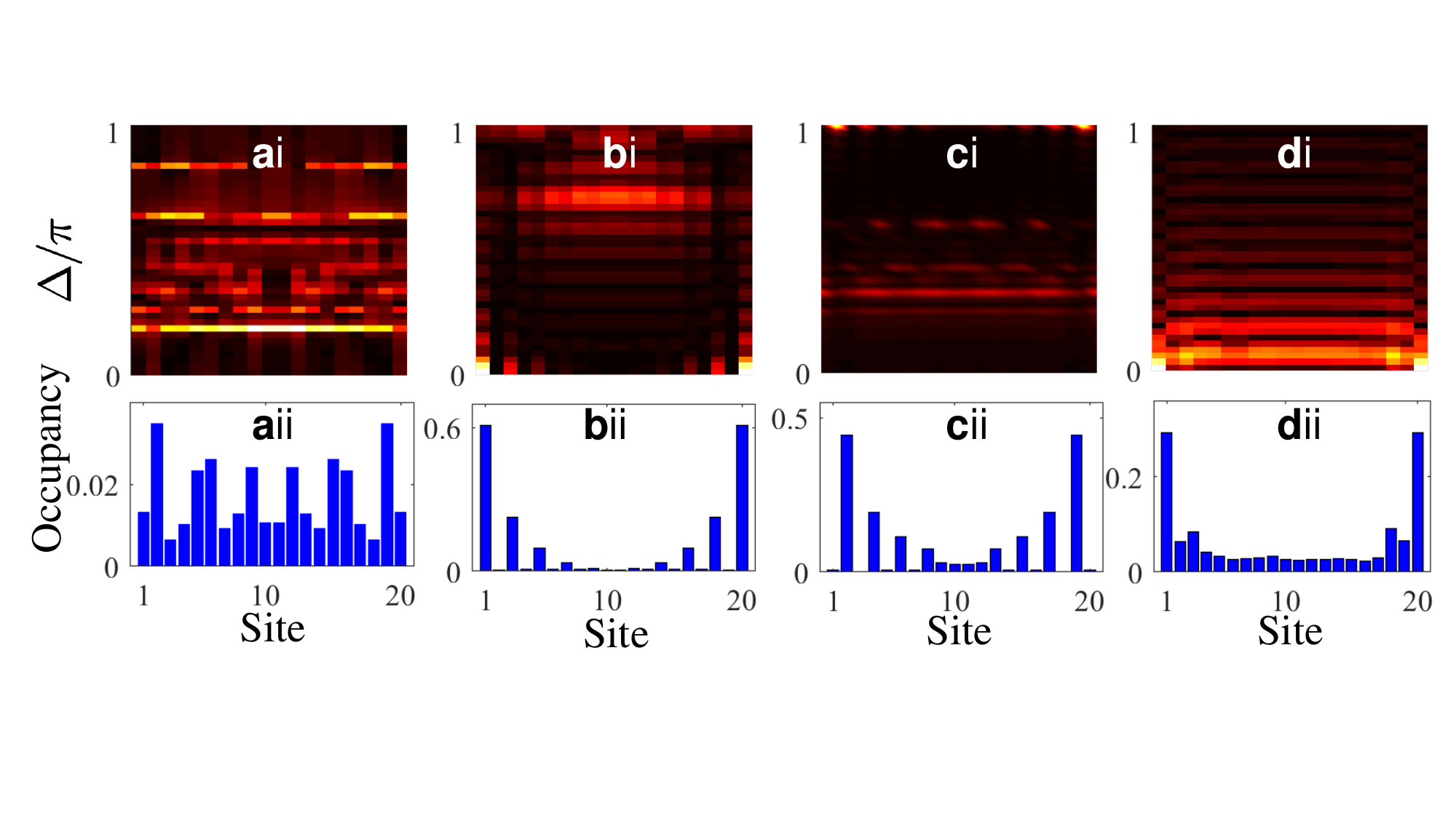}
	\caption{\textbf{Microwave detection of the edge states.} Occupancy of each site excited by the microwave field with detuning $\Delta$ for \textbf{a}i-\textbf{d}i and the corresponding selected Occupancy for \textbf{a}ii-\textbf{d}ii at $\Delta=0$ for \textbf{a}ii, \textbf{b}ii and \textbf{d}ii, or $\Delta=\pi$ for \textbf{c}ii, respectively. We use  $\mathbf{R}=(-0.7,0,0.8)$, $\boldsymbol{f}=(0.1,1,1)$, $d=0$, $\Omega=0.05$, $t_m=10T$ in \textbf{a}; $\mathbf{R}=(-0.5,0,0.8)$, $\boldsymbol{f}=(0.7,1,1)$, $d=0$, $\Omega=0.05$, $t_m=30T$ in \textbf{b}; $\mathbf{R}=(-0.5,0,0.8)$, $\boldsymbol{f}=(0.35,1,1)$, $d=0$, $\Omega=0.05$, $t_m=12T$ in \textbf{c}; $\mathbf{R}=(-0.9,1.6,0.8)$, $\boldsymbol{f}=(0.5,1,1)$, $d=0.3$, $\Omega=0.05$, $t_m=60T$ in \textbf{d}. Here $\mathbf{R}$ is the initial distance, $\boldsymbol{f}$  is the driving parameter, $d$ is the structural disorder strength, $\Omega$ is the Rabi frequency of the microwave, and $t_m$ is the total evolution time.}
	\label{fig3}
\end{figure}

Next, we propose to detect the Floquet topological phases with microwave spectroscopy. To this end, a weak global microwave field with the Rabi frequency $\Omega$ and detuning $\Delta$ is shined on. Then the system is described by a Hamiltonian $\hat{H}_\Omega=\hat{H}+\Omega/2\sum_i(\hat{b}_i^\dagger+\hat{b}_i)-\Delta\sum\hat{b}_i^\dagger\hat{b}_i$. The atoms are initially prepared in a vacuum state in the s-level. After applying the microwave probe for some periods of time $t_m$, an excitation can be created only if an eigenstate  energy matches the detuning $\Delta$. In Fig. \ref{fig3}a-\ref{fig3}d, we display the site-resolved probability on the p-level at the final time for different configurations (trivial regular, topological 0-phase regular, topological $\pi/T$-phase regular, topological amorphous, respectively), which can be detected by the fluorescence imaging in the experiment. We can observe a clear signal of the edge mode at either zero or $\pi$ detuning only in the topological phases, while a uniform distribution in the trivial cases.
%\section{Many-body case}\label{sec4}

\noindent \textbf{Many-body case.}
We now proceed to address the many-body case. The Hamiltonian (\ref{eq_ham}) can be mapped to an XY spin model via Matsubara-Matsuda transformation, $\hat{H}(t)=\sum_{i<j}V_{ij}(t) (\sigma_i^+\sigma_j^- +\sigma_j^+\sigma_i^-)$, where $\sigma_{i}^{\pm}=(\sigma_i^x\pm i\sigma_i^y)/2$ with $\sigma_j^s$ ($s=x,y,z$) being the Pauli matrices at site i. Correspondingly $m_z=\sum_i \sigma_i^z/2$ is a conserved quantity. The subspace of the spin Hamiltonian with $m_z=0$ is equivalent to the hard-core bosonic Hamiltonian at half-filling. We note that the effective Hamiltonian containing longer-range hoppings, despite without an interaction term like $\sigma_i^z\sigma_j^z$, can not be cast to a free fermionic model by the inverse Jordan-Wigner transformation, showing the many-body aspect of our model.

\begin{figure}[htbp]
	\centering
	\includegraphics[width=\textwidth]{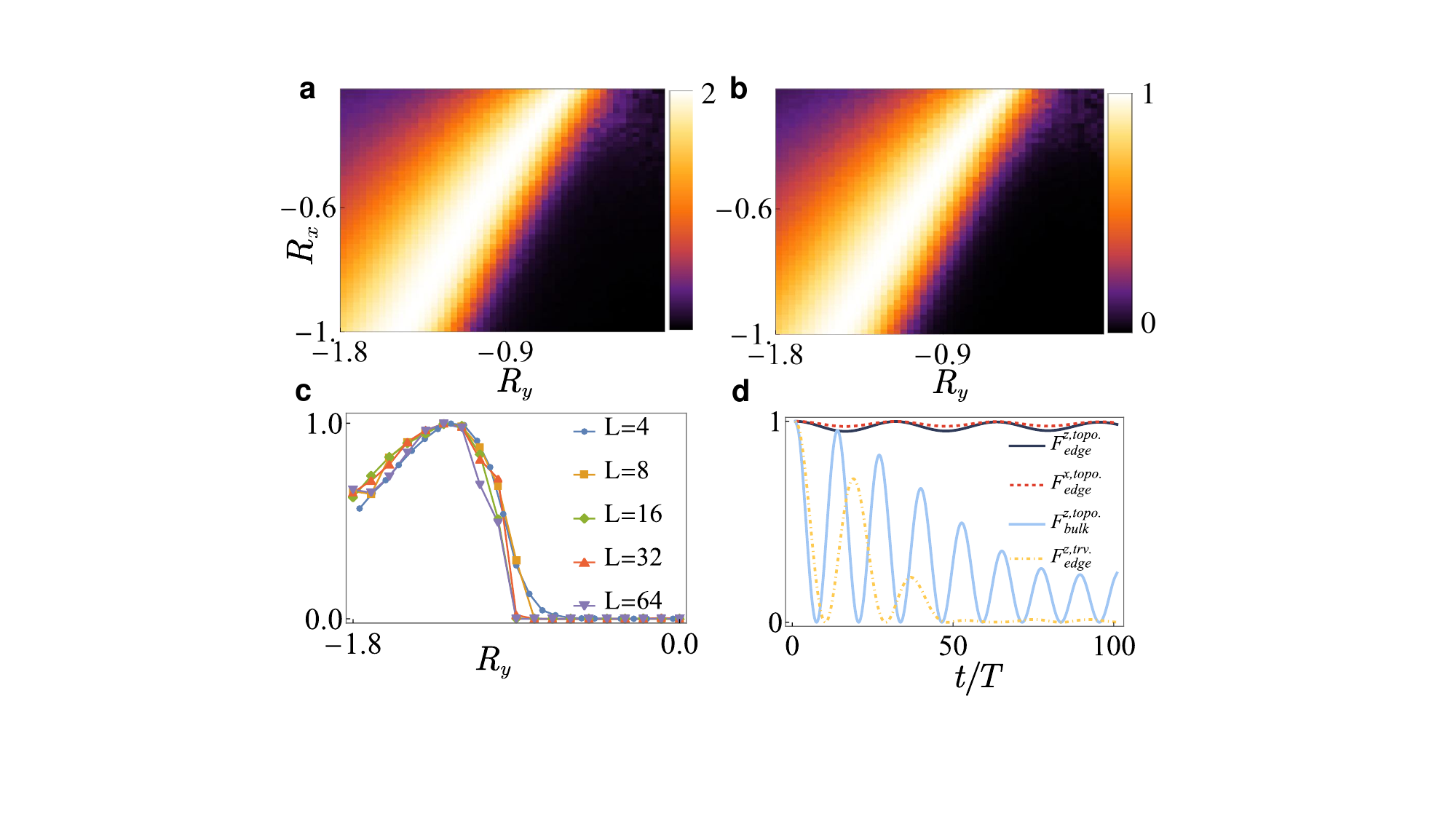}
	\caption{\textbf{Topological phase diagrams for the many-body case.} \textbf{a} Phase diagram characterized by topological entanglement entropy $S_E/{\rm ln}2$ (with the color shows the values) for a half-filled chain with $2L=8$ sites, and driving time duration $T_1=T_2=0.4$, driving paramters $\boldsymbol{f}=(0.8,1,1)$, initial distance $R_z=0.55$, structural disorder strength $d=0.8$, and averaged over 100 random configurations. \textbf{b} Phase diagram characterized by string order (with the color shows the values), with same parameter setting. \textbf{c} String order for systems with different lattice size. We choose the section of $R_x=-0.8$ here. \textbf{d} The trace fidelity for the topological case and trivial case for a lattice with $2L=12$ sites.}
	\label{fig4}
\end{figure}

We identify the topology of the many-body system by the topological entanglement entropy \cite{BZeng2015,BZeng2016,HLing2023},
\begin{equation}
S_E=S_{L/2}+S_{\tilde{L}/2}-S_{L/2 \cup \tilde{L}/2}-S_{L/2 \cap \tilde{L}/2},
\end{equation}
where $S_{L/2}$ ($S_{\tilde{L}/2}$) is the half-chain entanglement entropy of the reduced density matrix for the left half (second quarter and fourth quarter) part of the chain. The phase diagram according to $S_E$ is shown in Fig. \ref{fig4}a. The TEE is calculated for a Floquet eigenstate (analogous to the ground state of a static system) of a second order effective Hamiltonian, which is not degenerate for system with a periodic boundary. For the many-body case, the quasi-energy will increase when more excitations are created, $\varepsilon \sim \mathcal{O}(N)$. Therefore, there exists larger parameter regimes with $\rm{max}(\varepsilon) T<2\pi$ than the single particle case, where we could apply the Magnus expansion to simplify the numerical calculations (see Supplementary Note 4). The TEE of an SPT states yields a quantized value $2\ln 2$, due to that the topology is essentially carried on the boundary. In contrast, TEE vanishes in the trivial phase. The numerical results show a robust region of the topological phase under structural disorder for the many body case. In addition, as an measurable signature of the Floquet SPT, we calculate the string order $C^z_{\text {string }}=(-1)^{N-1} \langle\prod_{i=2}^{2 L-1} \sigma_i^z \rangle$ \cite{denNijs1989,Kennedy1992,Hida1992}, in Fig. \ref{fig4}b. The bulk of the topological states also acquire a finite long-range string order, and nearly vanishes in the trivial phase. We find that the TEE and the string order  exhibit almost the same behaviors. In these calculations, We use ED for $L<10$ and DMRG for $L>10$, and compare the results for different system sizes in Fig. \ref{fig4}c. The DMRG calculations are performed using  ITensor library \cite{itensor2022}.

To further diagnose the boundary physics, we calculate the edge fidelity, $F^\alpha_i=\mathrm{Tr}[\sigma_i^\alpha(t)\sigma_i^\alpha(0)]/N_{\mathrm{dim}}$ with $N_{\mathrm{dim}}$ ($\alpha=x,y,z$) being the dimension of the Hilbert space, following the definition in Ref. \cite{Potirniche2017}. The Floquet eigenstate is fourfold degenerate on open chains, as the edge modes for two sides can be either empty or occupied. As illustrated in Fig. \ref{fig4}d, the edge sites exhibits much longer coherence time than the bulk sites only in the topological regime, but quickly damps in the trivial phase. The long edge coherence of both $F^z$ and $F^x$  reveals the physical consequence of the SPT order in our model, and provides experimentally observable evidence.

\noindent \textbf{Discussion}

The topologically protected $\pi/T$ quasienergy excitations in the Floquet system always comes up in pairs, results in the subharmonic response at the edge \cite{Potirniche2017}. We show that our system exhibits time cystal behaviors at the edge sites, and such feature even exists under structural disorder (for details, see Supplementary Note 5). Although focusing here on the Rydberg atoms, our model may be generalized to other artificial quantum systems, such as trapped ion chain \cite{Bomantara2021,SLZhu2006,Blatt2012,JZhang2017,SChoi2017}, and superconducting qubits \cite{CYing2022,XMi2022}; and other symmetry class in higher dimensions.

In summary, we have proposed a concept of the Floquet amorphous topological matter. We have studied the topological properties in both single-particle and many-body levels. With various numerical calculations and careful comparison of different methods, the existence of rich Floquet SPT phases has been confirmed. We have found that diverse exotic SPT phases can be induced from the trivial static system by the periodic driving, or from the regular system by the disorder, along with localization transition. We have further addressed possible detection methods based on currently feasible technologies in the experiments of Rydberg atoms. Therefore, our work would provide a promising platform for
exploring the exotic physics of nonequilibrium systems that are elusive in nature.

\quad

\noindent \textbf{Data availability}

\noindent The data used to create the figures are available at Supplementary Data 1-4.

\quad

\noindent \textbf{Code availability}

\noindent All relevant codes are available from the corresponding authors upon reasonable request.

\quad

\noindent \textbf{Contributions}

\noindent P.H. and H.W. developed the ideas and designed the research, as well as performed analytical calculations and computational simulations. J.X.L. performed the DMRG simulations. Z.D.W. supervised the project. All authors discussed the results and contributed to the text of the manuscript.\\

\noindent \textbf{Competing interests}

\noindent The author declares no competing interests.

\quad

\acknowledgments

 This work was supported by the NSFC/RGC JRS grant (Grant No. N\_HKU 774/21), the CRF (Grant No. C6009-20G), GRF (Grants No. 17310622 and No. 17303023) of Hong Kong, National Natural Science Foundation (Grants No. 12405007), and Funds for Young Scientists of Chongqing Municipal Education Commission (Grants No.KJQN20240).
%\begin{appendix}
%\section{Pauli Matrix Representation}\label{appa}

%\end{appendix}

\begin{appendix}

\end{appendix}

\end{document}